# New expressions on the performance of a novel multi-hop relay-assisted hybrid FSO / RF communication system with receive diversity

Mohammad Ali Amirabadi

*Abstract*— **In this paper a novel comprehensive FSO / RF communication system is presented, in which a user is connected to the source Base Station via an RF link with receive diversity, then source and destination Base stations are connected via multi-hop hybrid parallel FSO / RF link. It is the first time that in a multi-hop FSO/RF system, receive diversity, signal selection at each hop, Demodulate & Forward relaying, Rayleigh fading in RF link, and wide range of atmospheric turbulences from moderate to strong are considered. Bit Error Rate and Outage Probability are investigated as performance criteria. New exact and asymptotic expressions are derived for these criteria, and MATLAB simulations are provided to verify them. In the results part, for the first time performance of such a structure is investigated at different number of receive antennas as well as relays; it's indicated that the proposed structure has low dependence on number of receive antennas, therefore single receive antenna scheme despite low complexity and power consumption, has the same performance as multi-antenna scheme. The proposed structure shows independent performance at moderate and strong atmospheric turbulence regimes. Hence, it does not require to adjust its parameters adaptively to maintain performance. According to these advantages, this structure is economically affordable and particularly suitable for urban communications which encounter frequent changes in atmospheric turbulence and require less complexity and power consumption.**

## I.  INTRODUCTION

**I**ntensity Modulation / Direct Detection (IM/DD) based on On-Off-Keying (OOK), because of simple implementation, is mostly used in FSO system [1]. In OOK detection threshold is adjusted based on Channel State Information (CSI), therefore the channel must be estimated. Pulse Position Modulation (PPM), is another modulation used in FSO systems, which has lower spectral efficiency than OOK, but fixed detection threshold. Subcarrier Intensity Modulation (SIM), due to its high spectral efficiency, is an appropriate alternative for PPM and OOK, but suffers from carrier frequency and phase synchronization.

Terrestrial FSO link, due to easy and low-cost installation, license-free spectrum, and high data rate and security, is a competitor for traditional RF system. FSO performance is strongly affected by the weather conditions; but even in clear weather, atmospheric turbulences, caused by temperature and pressure inhomogeneity in the atmosphere, mismatch it. Scintillation, one of the problems of atmospheric turbulence causes random fluctuations of the received signal intensity. Atmospheric turbulence can be modeled as Log-Normal [2], Gamma-Gamma [3], K [4], I-K, H-K [5], M [6], and Negative Exponential [7]; among them Gamma-Gamma and negative Exponential models, are in high accompany with experimental results obtained for moderate to strong and saturate regimes, respectively [8].

One of the main challenges in front of FSO system is misalignment of transceivers, known as pointing error effect, which significantly degrades performance of FSO system. Based on vertical and horizontal displacements of the transreceiver, this effect is divided to zero boresight and non-zero boresight effects. In zero boresight, horizontal and vertical displacements on the receiver plane are modeled by zero mean Gaussian distribution, whereas in non-zero boresight, horizontal and vertical displacements are modeled by non-zero mean Gaussian distribution. Radial displacements, in zero boresight error and non-zero boresight error are modeled by Rayleigh and Rician distributions, respectively. Aperture averaging is a low-cost, simple and useful method to compensate mitigation caused by pointing error [9, 10].

Relay-assisted is known as a solution of increasing the system capacity with low power requirement. Multi-hop FSO systems have advantages of both FSO and cooperative systems, such as high bandwidth, data rate, capacity, and better performance. The main difference between relay-assisted systems is related to the processing they make on the signal. In these systems, the received signal is either amplified [11] or decoded [12] or detected [13], and then forwarded. Each of the mentioned processing has its own advantages and due to consumer demands, such as power, accuracy, latency, complexity, and cost, one of them can be implemented. This is the first time that in a multi-hop relay assisted hybrid parallel FSO / RF system, the

M. A. Amirabadi is with the School of Electrical Engineering, Iran University of Science and Technology (IUST), Tehran 1684613114, Iran(email: m_amirabadi@elec.iust.ac.ir)

received signal is demodulated and forwarded at each hop. This processing takes place in two time slots; first received signal is demodulated and regenerated, then the regenerated signal is modulated and forwarded. The error occurrence in this protocol can be described as follows: symbol 1 is transmitted, if both relay and destination detect it wrongly, the error is compensated and so the whole system remains without error, else, an error occurs.

Besides advantages of FSO, its disadvantages such as high sensitivity to atmospheric turbulence and pointing error, severely limits its practical applications. One solution is to combine FSO and RF systems [14]. The so-called hybrid FSO / RF links are highly reliable, accessible and provide high capacity and data rate. FSO/RF papers are either single-hop, or dual-hop or multi-hop. Single-hop structures hybrid parallel FSO/RF link [15-19]. So compared with them, novelties of this work are multi-hop, receive diversity, demodulate and forward relaying, and mathematical solution. A new paper has investigated use of diversity in single-hop structures for the first time [20]. It has used MRC and EGC combiners at the FSO and RF receivers of parallel FSO/RF link. So compared with it, novelties of this work are implementation of Selection Combiner in RF link, plus the other things mentioned for single hop. Dual hop papers mostly use series FSO and RF structure with RF at first and FSO at second link [12, 13, 21, 22]. [23] used a backup FSO link between source and destination. So compared with them novelties of this work are implementation of parallel FSO/RF link at each hop, plus the other things mentioned for single hop. Some papers used multi-user scheme [24]. There are few works published on multi-hop FSO/RF system [25-28], which all of them have investigated Outage Probability. In multi-hop structures, hops can be serial [27], or parallel [28].

In multi-hop systems, one user acts as the transmitter and the other users can act as relays, when the received SNR, at the destination, reduces to a threshold level, other users start relaying data. In fact users together form an array of distributed antennas that can achieve diversity and multiplexing gains as MIMO systems [29]. Relaying information through a multi-hop link reduces the total power consumption. Thereby, the battery may storage its charge longer [30].

Space Diversity is an efficient way of compensating the mitigation caused by atmospheric turbulences. To the best of the author's knowledge, it is the first time that in a multi-hop hybrid parallel FSO / RF structure, receive diversity is used. In this technique, the receiver, by combining different received copies of the signal can better recover the original signal. There are several combiners for this purpose, such as Maximum Ratio Combining (MRC), Equal Gain Combining (EGC), and Selection Combining (SC) [31].

In this paper, a novel multi-hop relay-assisted hybrid FSO / RF system with received diversity, connects the mobile user to the Base Stations. It is the first time that such a structure is investigated at wide range of atmospheric turbulences, from moderate to saturate regimes. At this structure, data is demodulated and forwarded simultaneously through parallel FSO / RF link. FSO link at moderate to strong regime is

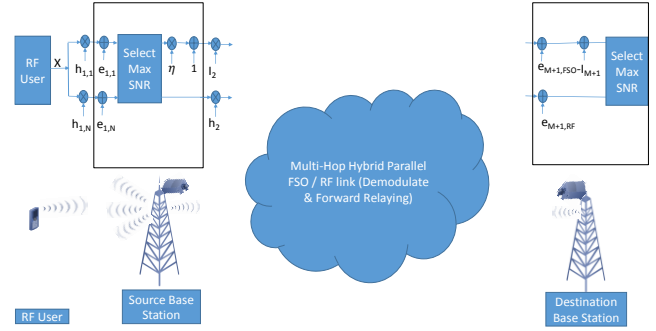

Fig. 1. The proposed multi-hop relay-assisted hybrid FSO / RF system with described by Gamma-Gamma atmospheric turbulence with the effect of pointing error and at saturate regime is described by Negative Exponential atmospheric turbulence, and RF link has Rayleigh fading. For the first time, new exact and asymptotic expressions are derived in closed-form for BER and $P_{out}$ of the proposed structure. MATLAB simulations are provided in order to validate the obtained results. Use of hybrid parallel FSO / RF link, despite more power consumption, significantly improves performance, reliability and accessibility of the link; also multi-hop relaying significantly increases capacity and reduces total power consumption.

The remainder of this study is organized as follows: section 2 describes system model. In sections 3 and 4, Outage Probability and Bit Error rate of the proposed structure are discussed, respectively. Section 5 provides simulation results and discussions. Section 6 is the conclusion.

## II. SYSTEM MODEL

As can be seen in Fig. 1, $x$ is the transmitted signal from RF user, the received signal at the first relay (source Base Station) with multiple receive antenna becomes as follow:

$$y_{1,i} = h_{1,i}x + e_{1,i}, \tag{1}$$

where, $e_{1,i}; i = 1,2,...,N$ is the Additive White Gaussian Noise (AWGN), with zero mean and $\sigma_{RF}^2$ variance, at the first relay (source Base Station) input, and $h_i$ is the fading coefficient of $i-th$ path between user and first relay. At the first relay selection combining scheme is used, therefore among received RF signals, one with the highest SNR is demodulated and regenerated. One copy of regenerated signal is converted to optical signal by conversion efficiency of $\eta$, and a DC bias with unit amplitude is added to it (in order to FSO signal be positive), and then modulated and forwarded via FSO link, another copy of regenerated signal is modulated and forwarded in RF link. At the other relays, between received FSO and RF signals, one with higher SNR is demodulated, regenerated and forwarded trough FSO and RF links.

In this paper, in order to have a comprehensive investigation on the proposed structure, wide range of atmospheric turbulence regimes, from moderate to saturate, are considered. Although there are many model suitable, but the best models preserving this end are Gamma-Gamma at moderate to strong and Negative Exponential at saturate regime. Also it is the first time that RF link in a multi-hop FSO/RF structure is investigated at Rayleigh fading. The probability density function (pdf) and Cumulative Distribution Function (CDF) of Gamma-Gamma distribution with the effect of pointing error [15,32], the CDF of Negative Exponential and Rayleigh distributions are respectively as follows:

$$f_\gamma(\gamma) = \frac{\xi^2}{2\Gamma(\alpha)\Gamma(\beta)\gamma} G_{1,4}^{3,0}\left(\alpha\beta\kappa\sqrt{\frac{\gamma}{\bar{\gamma}_{FSO}}} \middle| \begin{matrix} \xi^2+1 \\ \xi^2, \alpha, \beta \end{matrix}\right), \quad (2)$$

$$F_\gamma(\gamma) = \frac{\xi^2}{\Gamma(\alpha)\Gamma(\beta)} G_{2,4}^{3,1}\left(\alpha\beta\kappa\sqrt{\frac{\gamma}{\bar{\gamma}_{FSO}}} \middle| \begin{matrix} 1, \xi^2+1 \\ \xi^2, \alpha, \beta, 0 \end{matrix}\right), \quad (3)$$

$$F_\gamma(\gamma) = 1 - e^{-\lambda\sqrt{\frac{\gamma}{\bar{\gamma}_{FSO}}}}, \quad (4)$$

$$F_\gamma(\gamma) = 1 - e^{-\frac{\gamma}{\bar{\gamma}_{RF}}}, \quad (5)$$

where $G_-^-\left(. \middle| \begin{matrix} - \\ - \end{matrix}\right)$ is the Meijer-G function [33], $\alpha, \beta$ are Gamma-Gamma atmospheric turbulence characterization parameters, $\xi$ is pointing error characterization parameter [32] and $\Gamma(.)$ is Gamma function [33]. $\bar{\gamma}_{FSO} = E[x^2]\eta^2/\sigma_{FSO}^2$ and $\bar{\gamma}_{RF} = E[x^2]/\sigma_{RF}^2$ are the average SNR at FSO and RF receiver input, respectively and $\sigma_{FSO}^2$ and $\sigma_{RF}^2$ are AWGN variances of input noise at FSO and RF receivers.

Assuming, independent and identically distributed RF paths, and according to the first relay selects signal with the highest SNR, the CDF of instantaneous SNR at the first relay input becomes as follow:

$$F_{\gamma_1}(\gamma) = \Pr\left(max(\gamma_{1,1}, \gamma_{1,2}, ..., \gamma_{1,N}) \le \gamma\right) = \\ \Pr(\gamma_{1,1} \le \gamma, \gamma_{1,2} \le \gamma, ..., \gamma_{1,N} \le \gamma) = \\ \prod_{i=1}^{N}\Pr(\gamma_{1,i} \le \gamma) = \prod_{i=1}^{N} F_{\gamma_{1,i}}(\gamma) = \left(1 - e^{-\frac{\gamma}{\bar{\gamma}_{RF}}}\right)^N. \quad (6)$$

where $\gamma_1$ is instantaneous SNR at the first relay input. Between received FSO and RF signals at $j - th; j = 2,3, ... M$ relay, one with higher SNR is selected; therefore, the CDF of instantaneous SNR at the $j - th$ relay input becomes equal to:

$$F_{\gamma_j}(\gamma) = \Pr(max(\gamma_{j,1}, \gamma_{j,2}) \le \gamma) = \Pr(\gamma_{j,1} \le \\ \gamma, \gamma_{j,2} \le \gamma) = F_{\gamma_{j,1}}(\gamma) F_{\gamma_{j,2}}(\gamma). \quad (7)$$

where $\gamma_j$ is instantaneous SNR at the $j - th$ relay input. The last equality is because of independence of FSO and RF links.

## III. OUTAGE PROBABILITY

In the proposed structure, outage occurs while the SNR of received signal at each relay comes down below a threshold level. Therefore, $P_{out}$ of the proposed structure, is calculated as follows:

$$P_{out}(\gamma_{th}) = \Pr\{\gamma_1, \gamma_2, ..., \gamma_{M+1}\} \le \gamma_{th}\} = 1 - \quad (8) \\ \Pr\{\gamma_1 \ge \gamma_{th}, \gamma_2 \ge \gamma_{th}, ..., \gamma_{M+1} \ge \gamma_{th}\} = 1 - \\ (1 - \Pr\{\gamma_1 \le \gamma_{th}\}) (1 - \Pr\{\gamma_2 \le \\ \gamma_{th}\}) (1 - \Pr\{\gamma_{M+1} \le \gamma_{th}\}) = 1 - \left(1 - \\ F_{\gamma_1}(\gamma_{th})\right)\left(1 - F_{\gamma_2}(\gamma_{th})\right)...\left(1 - F_{\gamma_{M+1}}(\gamma_{th})\right).$$

Which means the availability of the through system is multiplication of each hop availability. Except the first link, others are the same structures, so have the same availability. Accordingly, $P_{out}$ of the proposed structure becomes equal to:

$$P_{out}(\gamma_{th}) = 1 - \left(1 - F_{\gamma_1}(\gamma_{th})\right)\left(1 - F_{\gamma_2}(\gamma_{th})\right)^M. \quad (9)$$

According to (7), and by substituting (3), (5) and (6) into (9), $P_{out}$ of the proposed structure in Gamma-Gamma atmospheric turbulence with the effect of pointing error becomes as follows:

$$P_{out}(\gamma_{th}) = 1 - \left(1 - \left(1 - e^{-\frac{\gamma_{th}}{\bar{\gamma}_{RF}}}\right)^N\right)\left(1 - \quad (10) \\ \frac{\xi^2}{\Gamma(\alpha)\Gamma(\beta)}\left(1 - e^{-\frac{\gamma_{th}}{\bar{\gamma}_{RF}}}\right) G_{2,4}^{3,1}\left(\alpha\beta\kappa\sqrt{\frac{\gamma_{th}}{\bar{\gamma}_{FSO}}} \middle| \begin{matrix} 1, \xi^2+1 \\ \xi^2, \alpha, \beta, 0 \end{matrix}\right)\right)^M.$$

In derivation of BER, the above term will be used, but its somehow complex and needs to be simplified more. By substituting binomial expansion of $\left(1 - e^{-\frac{\gamma_{th}}{\bar{\gamma}_{RF}}}\right)^N$ and $\left(1 - \frac{\xi^2}{\Gamma(\alpha)\Gamma(\beta)}\left(1 - e^{-\frac{\gamma_{th}}{\bar{\gamma}_{RF}}}\right) G_{2,4}^{3,1}\left(\alpha\beta\kappa\sqrt{\frac{\gamma_{th}}{\bar{\gamma}_{FSO}}} \middle| \begin{matrix} 1, \xi^2+1 \\ \xi^2, \alpha, \beta, 0 \end{matrix}\right)\right)^M$, $P_{out}$ of the proposed structure in Gamma-Gamma atmospheric turbulence with the effect of pointing error is as follows:

$$P_{out}(\gamma_{th}) = 1 + \sum_{k=1}^{N}\sum_{t=0}^{M}\sum_{u=0}^{t}\Omega e^{-\frac{(k+u)\gamma_{th}}{\bar{\gamma}_{RF}}} \times \quad (11) \\ \left(\frac{\xi^2}{\Gamma(\alpha)\Gamma(\beta)} G_{2,4}^{3,1}\left(\alpha\beta\kappa\sqrt{\frac{\gamma_{th}}{\bar{\gamma}_{FSO}}} \middle| \begin{matrix} 1, \xi^2+1 \\ \xi^2, \alpha, \beta, 0 \end{matrix}\right)\right)^t,$$

where $\Omega = \binom{N}{k}\binom{M}{t}\binom{t}{u}(-1)^{k+t+u}$. As can be seen the weight of the CDF of RF link in the above expression is more and it's something logical because there RF links are more than FSO links.

According to (7), and by substituting (4), (7) and (6) into (9), and substituting binomial expansion of $\left(1 - e^{-\frac{\gamma_{th}}{\bar{\gamma}_{RF}}}\right)^N$ and $\left(1 - \left(1 - e^{-\frac{\gamma_{th}}{\bar{\gamma}_{RF}}}\right)\left(1 - e^{-\lambda\sqrt{\frac{\gamma_{th}}{\bar{\gamma}_{FSO}}}}\right)\right)^M$, $P_{out}$ of the proposed structure in Negative Exponential atmospheric turbulence becomes as follow:

$$P_{out}(\gamma_{th}) = 1 + \sum_{k=1}^{N}\sum_{t=0}^{M}\sum_{u=0}^{t}\sum_{v=0}^{t} \Lambda e^{-\frac{(k+u)\gamma\gamma_{th}}{\overline{\gamma}_{RF}}} e^{-\lambda v \sqrt{\frac{\gamma_{th}}{\overline{\gamma}_{FSO}}}}, \quad (12)$$

where $\Lambda = \binom{N}{k}\binom{M}{t}\binom{t}{u}\binom{t}{v}(-1)^{k+t+u+v}$. Although there are 3 and 4 summations in (11) and (12); but it should be considered that this complexity is because of assuming a complex structure.

## IV. BIT ERROR RATE

In this paper DPSK modulation is used for both FSO and RF systems. BER of DPSK modulation can be calculated analytically using the following formula [14]:

$$P_e = \frac{1}{2}\int_0^\infty e^{-\gamma} F_\gamma(\gamma)d\gamma = \frac{1}{2}\int_0^\infty e^{-\gamma} P_{out}(\gamma)d\gamma. \quad (13)$$

where the last inequality is because $F_\gamma(\gamma) = P_{out}(\gamma)$. BER of DPSK modulation in Gamma-Gamma atmospheric turbulence with the effect of pointing error can be obtained by substituting (11) into (13):

$$P_e = \frac{1}{2}\int_0^\infty e^{-\gamma}\left\{1 + \sum_{k=1}^{N}\sum_{t=0}^{M}\sum_{u=0}^{t} \Omega e^{-\frac{(k+u)\gamma}{\overline{\gamma}_{RF}}} \times \right. \quad (14)$$
$$\left. \left(\frac{\xi^2}{\Gamma(\alpha)\Gamma(\beta)} G_{2,4}^{3,1}\left(\alpha\beta\kappa\sqrt{\frac{\gamma}{\overline{\gamma}_{FSO}}} \left| \begin{matrix} 1, \xi^2+1 \\ \xi^2, \alpha, \beta, 0 \end{matrix}\right.\right)\right)^t\right\}d\gamma.$$

When $> 2$, because of multiplication of three Meijer-G and one exponential functions, the above integral cannot be solved. The main novelty of mathematical calculations of this paper is this section; which derives new asymptotic and exact expressions in closed-form for BER and outage probability of the proposed structure in Gamma-Gamma atmospheric turbulence with the effect of pointing error.

### A. Exact BER for Gamma-Gamma atmospheric turbulence with the effect on pointing error

By substituting exact equivalent expression for the CDF of Gamma-Gamma atmospheric turbulence with the effect of pointing error from Appendix A into (11), and by substituting binomial expansion of $\left(X_0\left(\frac{\gamma_{th}}{\overline{\gamma}_{FSO}}\right)^{\frac{\xi^2}{2}} + \sum_{n=0}^{\infty}Y_n\left(\frac{\gamma_{th}}{\overline{\gamma}_{FSO}}\right)^{\frac{n+\alpha}{2}} + \sum_{n=0}^{\infty}Z_n\left(\frac{\gamma_{th}}{\overline{\gamma}_{FSO}}\right)^{\frac{n+\beta}{2}}\right)^t$ as $\sum_{k_1=0}^{t}\sum_{k_2=0}^{k_1}\binom{t}{k_1}\binom{k_1}{k_2}\left(X_0\left(\frac{\gamma_{th}}{\overline{\gamma}_{FSO}}\right)^{\frac{\xi^2}{2}}\right)^{t-k_1} \times \left(\sum_{n=0}^{\infty}Y_n\left(\frac{\gamma_{th}}{\overline{\gamma}_{FSO}}\right)^{\frac{n+\alpha}{2}}\right)^{k_1-k_2}\left(\sum_{n=0}^{\infty}Z_n\left(\frac{\gamma_{th}}{\overline{\gamma}_{FSO}}\right)^{\frac{n+\beta}{2}}\right)^{k_2}$ and after some mathematical simplifications, $P_{out}$ of the proposed structure becomes equal to:

$$P_{out}(\gamma_{th}) = 1 +$$
$$\sum_{k=1}^{N}\sum_{t=0}^{M}\sum_{u=0}^{t}\sum_{k_1=0}^{t}\sum_{k_2=0}^{k_1}\sum_{n=0}^{\infty}\Omega\binom{t}{k_1}\binom{k_1}{k_2}X_0^{t-k_1} \times$$
$$\left(Y_n^{(k_1-k_2)} * Z_n^{(k_2)}\right) e^{-\frac{(k+u)\gamma_{th}}{\overline{\gamma}_{RF}}}\left(\frac{\gamma_{th}}{\overline{\gamma}_{FSO}}\right)^{\frac{n+\xi^2(t-k_1)+\alpha(k_1-k_2)+\beta k_2}{2}}. \quad (15)$$

By substituting (15) into (13), BER of DPSK modulation in Gamma-Gamma atmospheric turbulence with the effect of pointing error becomes equal to:

$$P_e = \frac{1}{2}\left\{1 + \sum_{k=1}^{N}\sum_{t=0}^{M}\sum_{u=0}^{t}\sum_{k_1=0}^{t}\sum_{k_2=0}^{k_1}\sum_{n=0}^{\infty}\Omega\binom{t}{k_1}\binom{k_1}{k_2} \times \right.$$
$$\left. X_0^{t-k_1}\left(Y_n^{(k_1-k_2)} * Z_n^{(k_2)}\right)\frac{(1/\overline{\gamma}_{FSO})^{\frac{n+\xi^2(t-k_1)+\alpha(k_1-k_2)+\beta k_2}{2}}}{\left(1+\frac{k+u}{\overline{\gamma}_{RF}}\right)^{1+\frac{n+\xi^2(t-k_1)+\alpha(k_1-k_2)+\beta k_2}{2}}}\right\}. \quad (16)$$

As can be seen in (15) and (16), N and M which are number of antennas and relays respectively, are above limits of the summation. According that these equations are not one-to-one, its not possible to derive mathematical support for the impact of $M$ and $N$ on the performance of the proposed multi-hop FSO/RF structure; but it's possible to do through simulations. This is done in this paper for the first time at the results section.

### B. Asymptotic BER of Gamma-Gamma atmospheric turbulence with the effect of pointing error

According that (15) and (16) are a bit complex and it's not easy to enough insight to them, in the following section new asymptotic expressions are derived in closed-form for BER and outage probability of the proposed structure in Gamma-Gamma atmospheric turbulence with the effect of pointing error.

By substituting CDF of Gamma-Gamma atmospheric turbulence with the effect of pointing error, from Appendix B into (11), $P_{out}$ of the proposed system becomes as follows:

$$P_{out}(\gamma_{th}) \cong$$
$$\begin{cases} 1 + \sum_{k=1}^{N}\sum_{t=0}^{M}\sum_{u=0}^{t} e^{-\frac{(k+u)\gamma_{th}}{\overline{\gamma}_{RF}}}\Omega(\varpi)^t\gamma_{th}^{\frac{\beta t}{2}} & (1) \\ 1 + \sum_{k=1}^{N}\sum_{t=0}^{M}\sum_{u=0}^{t} e^{-\frac{(k+u)\gamma_{th}}{\overline{\gamma}_{RF}}}\Omega(\rho)^t\gamma_{th}^{\frac{\xi^2 t}{2}} & (2) \\ 1 + \sum_{k=1}^{N}\sum_{t=0}^{M}\sum_{u=0}^{t} e^{-\frac{(k+u)\gamma_{th}}{\overline{\gamma}_{RF}}}\Omega(\vartheta)^t\gamma_{th}^{\frac{\alpha t}{2}} & (3) \end{cases}. \quad (17)$$

As can be seen (17) is a simple linear equation It is very easy to have enough insight to it. As is expected by increasing $N$, $P_{out}$ decreases because k (which is related to $N$), is at the exponent multiplied by a minus sign. Also increasing number of relays, as expected increases $P_{out}$ because t (which is related to $M$), is in the power of $\gamma_{th}$. Increasing $M$ increases number of decisions made on the signal and this leads to increase in $P_{out}$. By substituting the result into (14), BER of DPSK modulation becomes equal to:

$$
P_e \cong \begin{cases}
1 + \sum_{k=1}^N \sum_{t=0}^M \sum_{u=0}^t \dfrac{\Omega(\varpi)^t \Gamma\left(\frac{\beta t}{2}+1\right)}{\left(1+\frac{k+u}{\bar{\gamma}_{RF}}\right)^{\frac{\beta t}{2}+1}} & (1) \\[4mm]
1 + \sum_{k=1}^N \sum_{t=0}^M \sum_{u=0}^t \dfrac{\Omega(\rho)^t \Gamma\left(\frac{\xi^2 t}{2}+1\right)}{\left(1+\frac{k+u}{\bar{\gamma}_{RF}}\right)^{\frac{\xi^2 t}{2}+1}} & (2) \\[4mm]
1 + \sum_{k=1}^N \sum_{t=0}^M \sum_{u=0}^t \dfrac{\Omega(\vartheta)^t \Gamma\left(\frac{\alpha t}{2}+1\right)}{\left(1+\frac{k+u}{\bar{\gamma}_{RF}}\right)^{\frac{\alpha t}{2}+1}} & (3)
\end{cases}
\tag{18}
$$

Also (18) is a linear function and simple insights of $P_{out}$ are also valid here with different reasons. But there is a beautiful insight here; amounts of k and t (which are related to $N$ and $M$, respectively), compared with $\bar{\gamma}_{RF}$ are very small and negligible. In the results section, it is shown that different number of antennas, especially at high $\bar{\gamma}_{RF}$ have the same effect on the performance of the proposed structure; and that's what (17) and (18) say.

Substituting (12) into (13), and substituting Meijer-G equivalent of $e^{-\lambda v \sqrt{\frac{\gamma}{\bar{\gamma}_{FSO}}}}$ as $\frac{1}{\sqrt{\pi}} G_{0,2}^{2,0}\left(\frac{(\lambda v)^2 \gamma}{4 \bar{\gamma}_{FSO}} \Big|_{\,0,0.5}^{\,-}\right)$ and using [33], BER of DPSK modulation in Negative Exponential atmospheric turbulence becomes equal:

$$
P_e = \frac{1}{2}\left\{ \sum_{k=1}^N \sum_{t=0}^M \sum_{u=0}^t \sum_{v=0}^t \Lambda \frac{1}{\sqrt{\pi}} \frac{1}{1+\frac{k+u}{\bar{\gamma}_{RF}}} \times \right.
$$
$$
\left. G_{1,2}^{2,1}\left( \frac{(\lambda v)^2}{4 \bar{\gamma}_{FSO}\left(1+\frac{k+u}{\bar{\gamma}_{RF}}\right)} \Big|_{\,0,0.5}^{\,0} \right) \right\}.
\tag{19}
$$

It is correct that with modern computing tools, it is now easier to produce analytical expressions. But it should be considered that in fact even modern computing tools cannot get forms shorter than formulations derived in this paper; because Meijer-G function is the shortest possible closed-form expression one can find for any kind of mathematical expressions, which used as a tool in many of the well-cited papers published by skilled authors in FSO performance investigation.

Although the mathematical formulations are a bit complex but it's not necessary to challenge with them in mathematical area, physical insights can be easily obtained by plotting their figures. The Meijer-G function have complex structure by itself, the proposed structure is also complex, it's reasonable that derived expressions become complex; isn't it? There are many papers in FSO system performance that used Meijer-G, without significant insight over mathematical expressions. They have just used math as a tool of deriving expressions needed for performance evaluation. But instead they brought enough physical insights while handling the results section.

## V. COMPARISON OF SIMULATION AND ANALYTICAL RESULTS

In this section simulation and analytical results, obtained for performance evaluation of the proposed structure, are compared and analyzed. Moderate to saturate atmospheric turbulence regimes are considered. It is assumed that FSO and RF links have the same average SNR ($\gamma_{avg} = \bar{\gamma}_{FSO} = \bar{\gamma}_{RF}$) and $\eta = 1$. The proposed structure is investigated at various number of receive antennas and relays. $M$ is number of relays and $N$ is number of receive antennas. $\gamma_{th}$ is the outage threshold SNR.

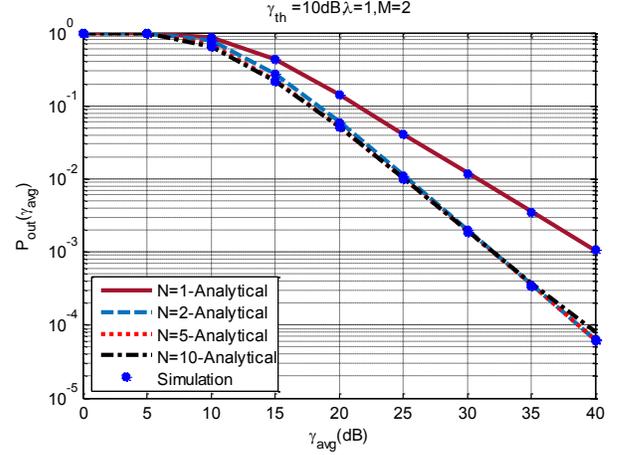

Fig. 2. Outage Probability of the proposed structure as a function of average SNR for various number of receive antennas for Negative Exponential atmospheric turbulence with unit variance, when number of relays is $M = 2$ and $\gamma_{th} = 10dB$.

In Fig. 2, Outage Probability of the proposed structure is plotted as a function of average SNR for various number of receive antennas for Negative Exponential atmospheric turbulence with unit variance, when number of relays is $M = 2$ and $\gamma_{th} = 10dB$. As can be seen, performance of the proposed structure has low dependence on number of receive antennas. Generally, a system with multiple receive antennas performs better than a system with single receive antenna. The proposed structure uses selection combiner and therefore, by increase of number of receive antenna, it is more likely that the SNR of the selected signal be higher than the threshold level.

In Fig. 3, Outage Probability of the proposed structure is plotted as a function of average SNR for various number of relays for Negative Exponential atmospheric turbulence with unit variance, when number of receive antennas is $N = 2$ and $\gamma_{th} = 10dB$. At $P_{out} = 10^{-4}$, $\gamma_{avg}$ difference between system performance at $M = 1$ and $M = 2,3,4$, is about $2dB$, $3dB$ and $4dB$, respectively. At wide range of target $P_{out}$, the same $\gamma_{avg}$ difference values can be observed. In series relay structure, performance degrades by relay addition. However, because $\gamma_{avg}$ difference values at different target $P_{out}$ are the same, only constant fraction of consumed power should be added to compensate this degradation, and it is not required additional processing to adjust this fraction adaptively.

In Fig. 4, Outage Probability of the proposed structure is plotted as a function of average SNR for various variances of Negative Exponential atmospheric turbulence, when number of users is $N = 2$, number of relays is $M = 2$ and $\gamma_{th} = 10dB$. It can be seen that the proposed structure is severely dependent on atmospheric turbulence variance; for example, at $P_{out} = 10^{-3}$, $\gamma_{avg}$ difference between system performance of the cases of $\lambda = 1$ and $\lambda = 2$, also between the cases of $\lambda = 1$ and $\lambda = 5$, is about $2dB$ and $5dB$, respectively. This difference changes at different target $P_{out}$, thereby, the proposed structure is not recommended for non-urban cells with high changes in atmospheric turbulence. But generally speaking at these places, because of the dependence on atmospheric turbulences, much more power is required to compensate the degradation caused by atmospheric turbulence.

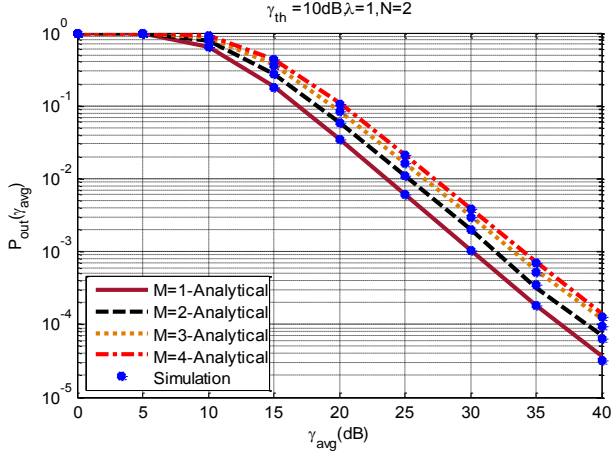

Fig. 3, Outage Probability of the proposed structure as a function of average SNR for various number of relays for Negative Exponential atmospheric turbulence with unit variance, when number of receive antenna is $N = 2$ and $\gamma_{th} = 10dB$.

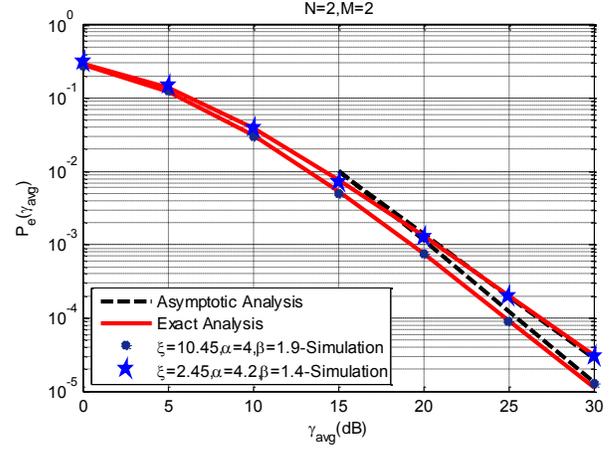

Fig. 5. Bit Error Rate of the proposed structure as a function of average SNR for moderate ($\alpha = 4, \beta = 1.9, \xi = 10.45$) and strong ($\alpha = 4.2, \beta = 1.4, \xi = 2.45$), regimes of Gamma-Gamma atmospheric turbulence with the effect of pointing error, when number of receive antenna is $N = 2$ and number of relays is $M = 2$.

In Fig. 5, Bit Error Rate of the proposed structure is plotted as a function of average SNR for moderate ($\alpha = 4, \beta = 1.9, \xi = 10.45$) and strong ($\alpha = 4.2, \beta = 1.4, \xi = 2.45$), regimes of Gamma-Gamma atmospheric turbulence with the effect of pointing error, when number of receive antenna is $N = 2$ and number of relays is $M = 2$. It can be seen that there is little $\gamma_{avg}$ difference between system performance at moderate and strong atmospheric turbulence regimes, for example, at $P_e = 10^{-4}$ and $P_e = 10^{-3}$, this difference is about $1.5dB$ and $2dB$, respectively. Therefore, this structure is suitable for urban cells which encounter with frequent changes in atmospheric turbulence. This structure is not sensitive to changes in atmospheric turbulence, therefore it does not need an adaptive processor in order to maintain performance. From this point of view, it has less complexity and installation cost.

In Fig. 6, Outage Probability of the proposed structure is plotted as a function of average SNR for various number of relays for moderate ($\alpha = 4, \beta = 1.9, \xi = 10.45$) regime of Gamma-Gamma atmospheric turbulence with the effects of pointing error, when number of antennas is $N = 2$ and $\gamma_{th} = 10dB$. It can be seen that at $P_{out} = 10^{-4}$, $\gamma_{avg}$ difference between system performance at the case of $M = 1$ and cases of $M = 2,3,4$ are about 1.5dB, 2dB, and 3dB, respectively. Of course, the degradation caused by relay addition is compensable by consuming power, therefore the proposed structure is particularly suitable for communication cells which service huge number of users.

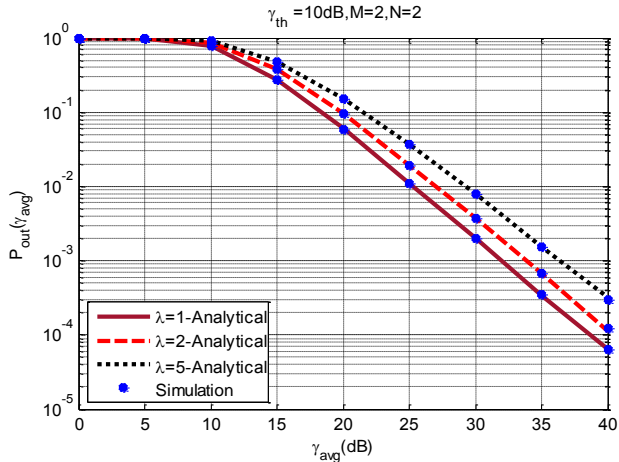

Fig. 4. Outage Probability of the proposed structure as a function of average SNR for various variances of Negative Exponential atmospheric turbulence, when number of users is $N = 2$, number of relays is $M = 2$ and $\gamma_{th} = 10dB$.

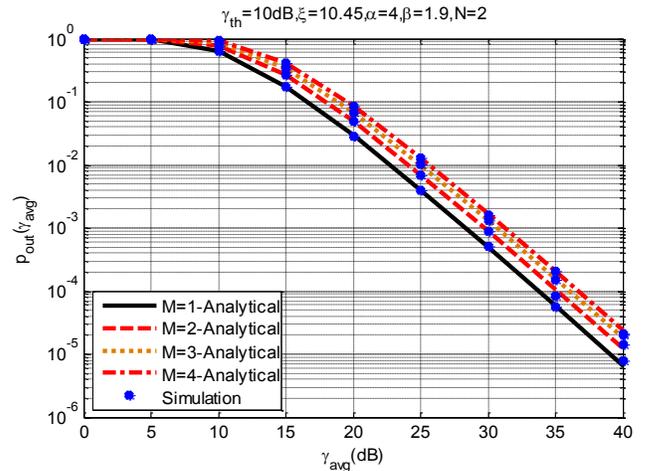

Fig. 6. Outage probability of the proposed structure as a function of average SNR for various number of relays for moderate ($\alpha = 4, \beta = 1.9, \xi = 10.45$) regime of Gamma-Gamma atmospheric turbulence with the effects of pointing error, when number of antennas is $N = 2$ and $\gamma_{th} = 10dB$.

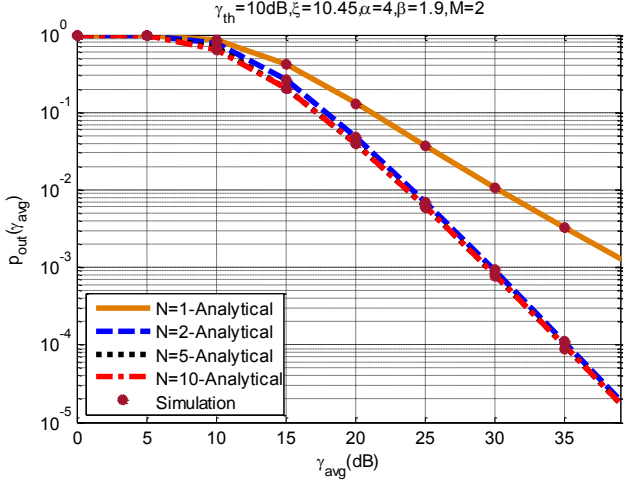

$\gamma_{th}=10dB, \xi=10.45, \alpha=4, \beta=1.9, M=2$

Fig. 7. Outage probability of the proposed structure as a function of average SNR for different number of receive antennas, for moderate ($\alpha = 4, \beta = 1.9, \xi = 10.45$) regime of Gamma-Gamma atmospheric turbulence with the effect of pointing error, when number of relays is $M = 2$ and $\gamma_{th} = 10dB$.

In Fig. 7, Outage probability of the proposed structure is plotted as a function of average SNR for different number of receive antennas, for moderate regime of Gamma-Gamma atmospheric turbulence with the effect of pointing error, when number of relay is $M = 2$ and $\gamma_{th} = 10dB$. As can be seen, the proposed structure has little dependence on the number of receive antennas. This is due to the selection of the antenna with the highest SNR. Since different received signal encounter independent fading, with the increase in the number of antennas, it is more likely to find signal with favorable SNR level.

## VI. CONCLUSION

In this study, a novel multi-hop relay-assisted hybrid FSO / RF structure is presented in which a user is connected to the source Base Station via RF link with receive diversity source and destination Base Stations are connected via a multi-hop hybrid FSO / RF link. Demodulate and forward protocol is used in this structure. The proposed structure has little dependence on the number of receive antennas, therefore, can use one receive antenna and eliminate the selection combiner while maintaining performance, hence cost, complexity, and power consumption reduces greatly. Also the difference between system performance at various number of relays is constant for wide range of target $P_{out}$, therefore, it does not require an adaptive processing in order to define the fraction of additional power for maintaining system performance, hence the proposed structure increases capacity while maintaining performance of the system, with low additional complexity and latency. The proposed structure is a bit more sensitive to Negative Exponential atmospheric turbulence than Gamma-Gamma atmospheric turbulence with the effect of pointing error. Hence the proposed structure is particularly suitable for urban communications which encounter frequent changes in atmospheric turbulence.

## APPENDIX A

Using [33], the pdf of Gamma-Gamma atmospheric turbulence with the effect of pointing error, becomes as follows:

$$f_{\gamma_{FSO}}(\gamma) = \frac{\xi^2 \Gamma(\alpha-\xi^2)\Gamma(\beta-\xi^2)}{2\Gamma(\alpha)\Gamma(\beta)\gamma}\left(\alpha\beta\kappa\sqrt{\frac{\gamma}{\bar{\gamma}_{FSO}}}\right)^{\xi^2} \times \quad (20)$$

$${}_1F_2\left(0;1-\alpha+\xi^2,1-\beta+\xi^2;\alpha\beta\kappa\sqrt{\frac{\gamma}{\bar{\gamma}_{FSO}}}\right) +$$

$$\frac{\xi^2 \Gamma(\xi^2-\alpha)\Gamma(\beta-\alpha)}{2\Gamma(\alpha)\Gamma(\beta)\Gamma(\xi^2+1-\alpha)\gamma}\left(\alpha\beta\kappa\sqrt{\frac{\gamma}{\bar{\gamma}_{FSO}}}\right)^{\alpha} \times$$

$${}_1F_2\left(\alpha-\xi^2;1-\xi^2+\alpha,1-\beta+\alpha;\alpha\beta\kappa\sqrt{\frac{\gamma}{\bar{\gamma}_{FSO}}}\right) +$$

$$\frac{\xi^2 \Gamma(\alpha-\beta)\Gamma(\xi^2-\beta)}{2\Gamma(\alpha)\Gamma(\beta)\Gamma(\xi^2+1-\beta)\gamma}\left(\alpha\beta\kappa\sqrt{\frac{\gamma}{\bar{\gamma}_{FSO}}}\right)^{\beta} \times$$

$${}_1F_2\left(\beta-\xi^2;1-\xi^2+\beta,1-\alpha+\beta;\alpha\beta\kappa\sqrt{\frac{\gamma}{\bar{\gamma}_{FSO}}}\right).$$

where ${}_pF_q(a_1,...,a_p;b_1,...,b_q;z)$ is the Hyper-Geometric function [33]. Using [33], the above expression becomes as follows:

$$f_{\gamma_{FSO}}(\gamma) = \frac{\xi^2 \Gamma(\alpha-\xi^2)\Gamma(\beta-\xi^2)}{2\Gamma(\alpha)\Gamma(\beta)}(\alpha\beta\kappa)^{\xi^2}\left(\frac{\gamma}{\bar{\gamma}_{FSO}}\right)^{\frac{\xi^2}{2}-1} +$$

$$\sum_{n=0}^{\infty}\frac{\xi^2 \Gamma(\beta-\alpha)(\alpha-\xi^2)_n}{2\Gamma(\alpha)\Gamma(\beta)(\xi^2-\alpha)(1-\xi^2+\alpha)_n(1-\beta+\alpha)_n n!}(\alpha\beta\kappa)^{n+\alpha}\left(\frac{\gamma}{\bar{\gamma}_{FSO}}\right)^{\frac{n+\alpha}{2}-1}$$

$$\sum_{n=0}^{\infty}\frac{\xi^2 \Gamma(\alpha-\beta)(\beta-\xi^2)_n}{2\Gamma(\alpha)\Gamma(\beta)(\xi^2-\beta)(1-\xi^2+\beta)_n(1-\alpha+\beta)_n n!}(\alpha\beta\kappa)^{n+\beta}\left(\frac{\gamma}{\bar{\gamma}_{FSO}}\right)^{\frac{n+\beta}{2}-1}, \quad (21)$$

where $(.)_n$ is the well-known pochhammer symbol. Integrating the above equation, CDF of Gamma-Gamma atmospheric turbulence with the effect of pointing error becomes as follows:

$$F_{\gamma_{FSO}}(\gamma) = \frac{\Gamma(\alpha-\xi^2)\Gamma(\beta-\xi^2)}{\Gamma(\alpha)\Gamma(\beta)}(\alpha\beta\kappa)^{\xi^2}\left(\frac{\gamma}{\bar{\gamma}_{FSO}}\right)^{\frac{\xi^2}{2}} +$$

$$\sum_{n=0}^{\infty}\frac{\xi^2 \Gamma(\beta-\alpha)(\alpha-\xi^2)_n}{(n+\alpha)\Gamma(\alpha)\Gamma(\beta)(\xi^2-\alpha)(1-\xi^2+\alpha)_n(1-\beta+\alpha)_n n!}(\alpha\beta\kappa)^{n+\alpha} \times$$

$$\left(\frac{\gamma}{\bar{\gamma}_{FSO}}\right)^{\frac{n+\alpha}{2}} + \sum_{n=0}^{\infty}\frac{\xi^2 \Gamma(\alpha-\beta)(\beta-\xi^2)_n}{(n+\beta)\Gamma(\alpha)\Gamma(\beta)(\alpha-\beta)(1-\xi^2+\beta)_n(1-\alpha+\beta)_n n!} \times$$

$$(\alpha\beta\kappa)^{n+\beta}\left(\frac{\gamma}{\bar{\gamma}_{FSO}}\right)^{\frac{n+\beta}{2}} = X_0\left(\frac{\gamma}{\bar{\gamma}_{FSO}}\right)^{\frac{\xi^2}{2}} + \sum_{n=0}^{\infty}Y_n\left(\frac{\gamma}{\bar{\gamma}_{FSO}}\right)^{\frac{n+\alpha}{2}} +$$

$$\sum_{n=0}^{\infty}Z_n\left(\frac{\gamma}{\bar{\gamma}_{FSO}}\right)^{\frac{n+\beta}{2}} \quad (22)$$

## APPENDIX B

From approximation [33], at $\gamma \gg 1$, the pdf of Gamma-Gamma atmospheric turbulence with the effect of pointing error becomes as follows:

$$f_\gamma(\gamma) \cong \begin{cases} \dfrac{\xi^2 \Gamma(\alpha-\beta)}{2\Gamma(\alpha)\Gamma(\beta)(\xi^2-\beta)\gamma}\left(\alpha\beta\kappa\sqrt{\dfrac{\gamma}{\bar{\gamma}_{FSO}}}\right)^\beta & \xi^2 > \beta, \alpha > \beta \\[3mm] \dfrac{\xi^2 \Gamma(\beta-\xi^2)\Gamma(\alpha-\xi^2)}{2\Gamma(\alpha)\Gamma(\beta)\gamma}\left(\alpha\beta\kappa\sqrt{\dfrac{\gamma}{\bar{\gamma}_{FSO}}}\right)^{\xi^2} & \alpha > \xi^2, \beta > \xi^2 \\[3mm] \dfrac{\xi^2 \Gamma(\beta-\alpha)}{2\Gamma(\alpha)\Gamma(\beta)(\xi^2-\alpha)\gamma}\left(\alpha\beta\kappa\sqrt{\dfrac{\gamma}{\bar{\gamma}_{FSO}}}\right)^\alpha & \beta > \alpha, \xi^2 > \alpha \end{cases}$$

(23)

Integrating above equation, CDF of Gamma-Gamma atmospheric turbulence with the effect of pointing error is as:

$$F_\gamma(\gamma) \cong \qquad\qquad\qquad (24)$$

$$\begin{cases} \dfrac{\xi^2 \Gamma(\alpha-\beta)}{\Gamma(\alpha)\Gamma(\beta+1)(\xi^2-\beta)}\left(\alpha\beta\kappa\sqrt{\dfrac{1}{\bar{\gamma}_{FSO}}}\right)^\beta \gamma^{\frac{\beta}{2}} & \xi^2 > \beta, \alpha > \beta \\[3mm] \dfrac{\Gamma(\beta-\xi^2)\Gamma(\alpha-\xi^2)}{\Gamma(\alpha)\Gamma(\beta)}\left(\alpha\beta\kappa\sqrt{\dfrac{1}{\bar{\gamma}_{FSO}}}\right)^{\xi^2} \gamma^{\frac{\xi^2}{2}} & \alpha > \xi^2, \beta > \xi^2 = \\[3mm] \dfrac{\xi^2 \Gamma(\beta-\alpha)}{\Gamma(\alpha+1)\Gamma(\beta)(\xi^2-\alpha)}\left(\alpha\beta\kappa\sqrt{\dfrac{1}{\bar{\gamma}_{FSO}}}\right)^\alpha \gamma^{\frac{\alpha}{2}} & \beta > \alpha, \xi^2 > \alpha \end{cases}$$

$$\begin{cases} \varpi\gamma^{\frac{\beta}{2}} & (1) \\ \rho\gamma^{\frac{\xi^2}{2}} & (2) \\ \vartheta\gamma^{\frac{\alpha}{2}} & (3) \end{cases}$$

## REFERENCES

[1] M. T. Dabiri, S. M. S. Sadough, M. A. Khalighi, "FSO channel estimation for OOK modulation with APD receiver over atmospheric turbulence and pointing errors", Optics Communications, Vol.402, Pp.577-584, Nov. 2017

[2] F. J. Lopez-Martinez, G. Gomez, J. M. Garrido-Balsells, "Physical Layer Security in Free-Space Optical Communications", IEEE Photonics Journal, Vol. 7, No. 2, Apr. 2015

[3] H. Kaushal, V.K. Jain, S. Ka, Free Space Optical Communication, Springer India, Pp-60, 2017

[4] H. A. Fadhil, A. Amphawan, H. A. B. Shamsuddin, T. Hussein Abd, H. M.R. Al-Khafaji, S.A. Aljunid, N. Ahmed, "Optimization of free space optics parameters: An optimum solution for bad weather conditions", Journal of Optik, Vol. 124, No.19, Pp. 3969-3973, Oct.2013

[5] P. Puri, P. Garg, M. Aggarwal, "Analysis of spectrally efficient two-way relay assisted free space optical systems in atmospheric turbulence with path loss", International Journal of Communication Systems, Vol. 29, No. 1, Pp. 99–112, Jan, 2016

[6] W. Liu, W. Shi, J. Cao, Y. Lv, K. Yao, S. Wang, J. Wang, X. Chi, " Bit error rate analysis with real-time pointing errors correction in free space optical communication systems", Journal of Optik, Vol. 125, No. 1, Pp.324–328, Jan, 2014

[7] M. Uysal, C. Capsoni, Z. Ghassemlooy, A. Boucouvalas, E. Udvary, Optical Wireless Communications An Emerging Technology, Springer International Publishing Switzerland, Pp-180, 2016

[8] W. Gappmair, H. E. Nistazakis, "Subcarrier PSK Performance in Terrestrial FSO Links Impaired by Gamma-Gamma Fading, Pointing Errors, and Phase Noise", Journal of Lightwave Technology, Vol. 35, No. 9, May, 2017

[9] P. K. Sharma, A. Bansal, P. Garg, T. A. Tsiftsis, R. Barrios, "Relayed FSO Communication with Aperture Averaging Receivers and Misalignment Errors", IET Communications, Vol. 11, No. 1, Jan. 2017

[10] M. A. Amirabadi, V. T. Vakili, "A new optimization problem in FSO communication system", IEEE Communications Letters. Vol. 22, No. 7, Apr. 2018.

[11] D. Huu Ai, D. T. Quang, N. N. Nam, H. D. Trung, D. T. Tuan, N. X. Dung, "Capacity analysis of amplify-and-forward free-space optical communication systems over atmospheric turbulence channels", Seventh International Conference on Information Science and Technology (ICIST) , May. 2017

[12] N. Varshney, P. Puri, "Performance Analysis of Decode-and-Forward-Based Mixed MIMO-RF/FSO Cooperative Systems With Source Mobility and Imperfect CSI", Journal of Lightwave Technology, Vol. 35, No. 11, June. 2017

[13] H. Saidi, N. Tourki, N. Hamdi, "Performance analysis of PSK modulation in DF dual-hop hybrid RF/FSO system over gamma gamma channel", International Symposium on Signal, Image, Video and Communications (ISIVC), Nov. 2016

[14] T. Rakia, H. C. Yang, M.S. Alouini, F. Gebali, "Outage Analysis of Practical FSO/RF Hybrid System with Adaptive Combining," IEEE Communications Letters,Vol.19, No.8, Aug. 2015

[15] T. Rakia, H. C. Yang, F. Gebali, M.S. Alouini, "Power Adaptation Based on Truncated Channel Inversion for Hybrid FSO/RF Transmission With Adaptive Combining", IEEE Photonics Journal, Vol.7, No.4, Aug. 2015

[16] M. A. Amirabadi, V. T. Vakili, "On the Performance of a CSI Assisted Dual-Hop Asymmetric FSO/RF Communication System over Gamma-Gamma atmospheric turbulence considering the effect of pointing error', international Congress on Science and Engineering, Mar. 2018

[17] V. V. Mai, A. T. Pham, "Adaptive Multi-Rate Designs for Hybrid FSO/RF Systems over Fading Channels", Globecom Workshops (GC Wkshps), 2015, Pp. 469-474

[18] Z. Kolka, Z. Kincl, V. Biolkova, D. Biolek, "Hybrid FSO/RF Test Link", 4th International Congress on Ultra Modern Telecommunications and Control Systems and Workshops (ICUMT), 2012

[19] A. AbdulHussein, A. Oka, T. T. Nguyen, L. Lampe, "Rateless Coding for Hybrid Free-Space Optical and Radio-Frequency Communication", IEEE Transactions on Wireless Communications, 2010, 9 , (3)

[20] M. A. Amirabadi, V. T. Vakili, "A novel hybrid FSO/RF communication system with receive diversity", arXiv preprint arXiv:1802.07348, Feb. 2018

[21] M. A. Amirabadi, V. T. Vakili, "Performance of a Relay-Assisted Hybrid FSO/RF Communication System", arXiv preprint arXiv:1803.00711. Mar. 2018

[22] M. A. Amirabadi, V. T. Vakili, "Performance analysis of hybrid FSO/RF communication systems with Alamouti Coding or Antenna Selection", arXiv preprint arXiv:1802.07286, Feb. 2018

[23] E. Soleimani-Nasab, M. Uysal, "Generalized Performance Analysis of Mixed RF/FSO Cooperative Systems", IEEE Transactions on Wireless Communications, 2016, 15, (1), Pp. 714 - 727

[24] L. Chen, W. Wang, C. Zhang, "Multiuser Diversity Over Parallel and Hybrid FSO / RF Links and Its Performance Analysis", IEEE Photonics Journal, 2016, 8, (3), Pp. 1-9

[25] B. Makki, T. Svensson, M. B. Pearce, M. S. Alouini, "Performance analysis of RF-FSO multi-hop networks", IEEE Wireless Communications and Networking Conference (WCNC), Mar.2017

[26] B. Makki, T. Svensson, M. B. Pearce, M. S. Alouini, "On the Performance of Millimeter Wave-based RF-FSO Multi-hop and Mesh Networks", IEEE Transactions on Wireless Communications, Mar. 2017

[27] M. Najafi, V. Jamali, R. Schober, "Optimal Relay Selection for the Parallel Hybrid RF/FSO Relay Channel: Non-Buffer-Aided and Buffer-Aided Designs", IEEE Transactions on Communications, Vol. 65, No. 7, Pp. 2794-2810, July. 2017

[28] H. Kazemi, M. Uysal, F. Touati, H. Haas, "Outage Performance of Multi-Hop Hybrid FSO/RF Communication Systems", 4th International Workshop on Optical Wireless Communications (IWOW), Sept.2015

[29] Y. W. Peter Hong, W. J. Huang, C. C. J. Kuo, Cooperative Communications and Networking Technologies and System Design, Springer US, 2010

[30] A. Yahya, LTE-A Cellular Networks, Springer International Publishing, 2017

[31] N. I. Miridakis, T. A. Tsiftsis, "EGC Reception for FSO Systems Under Mixture-Gamma Fading Channels and Pointing Errors", IEEE Communications Letters, Vol. PP, No. 99, Feb. 2017

[32] M. R. Bhatnagar, Z. Ghassemlooy, "Performance Analysis of Gamma-Gamma Fading FSO MIMO Links with Pointing Errors", Journal of Lightwave Technology, Vol. 34, No. 9, Pp.2158-2169, May. 2016

[33] http://functions.wolfram.com/HypergeometricFunctions/